\documentclass[12pt]{article}
\usepackage{fullpage,authblk,cite}
\usepackage{amsmath,amsthm,graphicx,amssymb,verbatim,listings}
\usepackage{wasysym,color,enumitem,mathcomp}
\usepackage[T1]{fontenc}     
\usepackage{lmodern, tipa}   
\usepackage[ pdftex, plainpages = false, pdfpagelabels,
                 pdfpagelayout = useoutlines,
                 bookmarks,
                 bookmarksopen = true,
                 bookmarksnumbered = true,
                 breaklinks = true,
                 linktocpage=all,
                 pagebackref=false,
                 colorlinks = true,
                 linkcolor = Blue,
                 urlcolor  = blue,
                 citecolor = BrickRed,
                 anchorcolor = green,
                 hyperindex = true,
                 hyperfigures
                 ]{hyperref}
\usepackage[usenames, dvipsnames]{xcolor}
\usepackage{xifthen} 

\newcommand{\ket}[1]{{\ensuremath{\left| #1 \right\rangle}}}

\newcommand{\arxiv}[2][]{\ifthenelse{\isempty{#1}}{\href{http://arxiv.org/abs/#2}{{\tt arXiv:\allowbreak{}#2}}} {\href{http://arxiv.org/abs/#2}{{\tt arXiv:\allowbreak{}#2 [#1]}}}}

\newcommand{\booktitle}{\textsl}
\newcommand{\hrefdoi}[2]{\href{https://dx.doi.org/#1}{#2}}

\begin{document}
\title{Whose Probabilities? About What? A Reply to Khrennikov}
\author[$\dag$]{Blake C.\ Stacey}
\affil[$\dag$]{Physics Department, University of
    Massachusetts Boston\protect\\ 100 Morrissey Boulevard, Boston MA 02125, USA}

\date{\small\today}

\maketitle

\begin{abstract}
In a recent article, Khrennikov claims that a particular theorem about
agreement between quantum measurement results poses a problem for the
interpretation of quantum mechanics known as QBism. Considering the
basic setup of that theorem in light of the meaning that QBism gives
to probability shows that the claim is unfounded.  
\end{abstract}

Conversations about quantum foundations often hare off into the forest
of unproductivity. It is not atypical for Bob to mount a criticism
that attacks his own preconceptions more than it does the actual views
of Alice. With considerable regret~\cite{Fuchs:2021}, we report here
that Khrennikov's recent critique of QBism falls into this category.

Since QBrevity is the soul of wit, we will refer to the literature for
detailed treatments of QBism~\cite{Fuchs:2013, Fuchs:2013b,
  Fuchs:2016}, how it developed~\cite{Stacey:2019}, the way it
addresses interactions between agents~\cite{DeBrota:2020}, the
technical work it motivates~\cite{Stacey:2021, DeBrota:2021} and so
on, instead just quoting the summary from the FAQ~\cite{DeBrota:2018}:
QBism is
\begin{quote}
  an interpretation of quantum mechanics in which the ideas of
  \emph{agent} and \emph{experience} are fundamental. A ``quantum
  measurement'' is an act that an agent performs on the external
  world. A ``quantum state'' is an agent's encoding of her own
  personal expectations for what she might experience as a consequence
  of her actions. Moreover, each measurement outcome is a personal
  event, an experience specific to the agent who incites
  it. Subjective judgments thus comprise much of the quantum
  machinery, but the formalism of the theory establishes the standard
  to which agents should strive to hold their expectations, and that
  standard for the relations among beliefs is as objective as any
  other physical theory.
\end{quote}

Khrennikov bases his argument~\cite{Khrennikov:2023} upon a theorem of
Ozawa~\cite{Ozawa:2019}. In the scenario that Ozawa considers, there
is a system, an environment and two remote observers. The theorem is
posed in terms of various mathematical entities: self-adjoint
operators $A$, $M_1$ and $M_2$ that are to be ``observables'' in the
von Neumann sense, a state vector $\ket{\psi}$ for the system and
another $\ket{\xi}$ for the environment, and a one-parameter family of
unitaries $U(t)$ to represent time evolution. Now, to a QBist, all of
these have the same status as probabilities do in the interepretation
of probability theory that QBism adopts, the personalist Bayesianism
of Ramsey and de Finetti. According to this school of thought, a
probability for an event is nothing more nor less than a gambling
commitment, a valuation by a specific agent of how much that agent
would stake on that event occurring. Any choice of state vector,
``observable'' or quantum channel is likewise a personal valuation, a
gambling commitment made in a way that is well-adapted to physics
calculations. Any probability extracted from combining these
quantities is necessarily, just like any other probability in
personalist Bayesianism, the possession of the agent who commits to
it. So, there is no way to mix the ingredients $A$, $M_1$, $M_2$ and
so forth to arrive at a conclusion that the personal experiences of
two agents will always agree, or that they will always disagree, or
anything in between.

To say it another way, suppose that one proves a theorem saying that
the result of measurement $X$ will agree with the result of
measurement $Y$ with probability 1. The personalist Bayesian may
justifiably ask, ``Whose head does that probability-of-unity live
in?'' Those gamblers whose background mesh of beliefs meet the
conditions of the theorem should, to be internally self-consistent,
assign probability 1 as the theorem says, but gamblers coming from
different backgrounds are not so bound.

As the quote above notes, QBism involves two levels of personalism: In
addition to its interpretation of probability, measurement outcomes
are also taken to be personal. This move of tying the sample space to
the agent was made in order to achieve consistency in the treatment of
Wigner's Friend and related paradoxes~\cite{DeBrota:2020}. Khrennikov
argues against this level of personalism, seemingly by declaring it to
be false and thus deducing that it must be false. Khrennikov
elaborates on an argument of Ozawa that begins by saying, ``Suppose
that two remote observers, I and II, simultaneously measure the same
observable.'' (\emph{According to whom?!,} the reader should already
be asking.) The conclusion is that ``quantum mechanics predicts that
they always obtain the same outcome''; but these words do not follow
from the mathematics. Whether the equations or any of the quantities
within them can actually refer to a pair of ``two remote observers''
measuring ``the same observable'' is an assumption dependent upon
one's interpretation of quantum theory. In QBism, that is simply not
what the equations can be talking about. Advocates of other
interpretations may for their own reasons find the premises of Ozawa's
theorem an inadequate representation of what it claims to be modeling
(a system, two observers, some devices, a common environment). For a
QBist, the argument simply does not get off the ground.

Alice, the proverbial user of quantum theory, might under some
conditions make a probability-1 prediction that the results of two
different actions she can take will be the same. There is no great
drama in this; but this is all that Ozawa's theorem or anything in its
vein can possibly deliver. When Khrennikov declares on the basis of
Ozawa's theorem that ``it's unnatural to consider outcomes of quantum
observations as [an] agent's personal experiences,'' he is imposing the
conclusion he desires upon the mathematics.

Conversational ripostes to QBism, we have noticed, sometimes confuse
themselves by switching from first-person to third-person midway
through. An interlocutor might ask, ``How can probabilities be
subjective in a scenario where\ldots,'' then proceeding to lay out a
situation described, implicitly or explicitly, in terms of state
vectors, unitaries and projectors. The assumption, perhaps a deeply
tacit one, is that \emph{some} objective values of these \emph{must}
exist. ``If we assume that probabilities are objective,'' it boils
down to asking, ``how can you say they are subjective?'' Khrennikov's
line of argument is a species of this.

Khrennikov writes of ``the difference between the accurate, von
Neumann, and inaccurate, noisy, quantum observables which are
represented by PVMs and POVMs
respectively''~\cite{Khrennikov:2023}. To a QBist, this way of
thinking is just baggage from classical mechanics. QBism treats the
quantum formalism as an empirically-motivated addition to decision
theory, and thus, the careful and expensive manipulations that Alice
might do in a quantum-computing laboratory are on a continuum with the
actions that she might do in her daily life~\cite{Mermin:2014,
  Fuchs:2017}. So, there is no fundamental reason to privilege those
POVMs whose elements happen to be projectors onto orthonormal
vectors. (They are not even a rich enough set to provide quantum
counterparts for all the questions we could ask in classical
physics~\cite{Peres:1993}.) Remember, a QBist maintains that the
choice to associate a particular set of self-adjoint matrices to a
particular set of possible experiences is a personal valuation by the
agent who is contemplating those experiences. Choosing to ascribe a
von Neumann ``observable'' rather than a POVM of another kind means
committing to a particular set of conditional gambles. Language about
other POVMs being ``inaccurate'' or ``noisy'' loads the narrative with
imagery about quantum measurements merely revealing the values that
were waiting to be found, and if it must be employed, should be used
more judiciously~\cite{Appleby:2016}. As Kemble wrote, ``[W]e get into
a maze of contradictions as soon as we inject into quantum mechanics
such concepts carried over from the language and philosophy of our
scientific ancestors''~\cite{Kemble:1937}.

Khrennikov contrasts QBism with ``the Copenhagen interpretation'', a
term that is a pet peeve of at least one QBist, as it has a profusion
of possible meanings, most of them apocryphal. For example, it is
often conflated with the ``shut up and calculate'' attitude (a saying
that is itself habitually misattributed~\cite{Mermin:2004}), which is
not in the slightest an attitude that Bohr could have
entertained~\cite{Bohr:1962, Casimir:1963}. While a start has been
made at surveying how QBism differs from all the things that have been
called Copenhagen~\cite{Mermin:2019}, this historical and
philosophical investigation remains incomplete. A further study of the
contrast would be welcome, but such an effort needs to be clear on
what QBism actually says.

\end{document}